\documentclass[%
 reprint,
 amsmath,amssymb,
 aps,
prb,
]{revtex4-2}
\usepackage{amsmath}
\usepackage{graphicx}
\usepackage{dcolumn}
\usepackage{bm}
\usepackage{amsmath}
\usepackage{lineno}
 \usepackage{hyperref}
 \hypersetup{
     colorlinks=true,
     linkcolor=blue,
     filecolor=blue,
     citecolor = blue,      
     urlcolor=blue,
     }

\usepackage{tikz,xcolor,hyperref}

\definecolor{lime}{HTML}{A6CE39}
\DeclareRobustCommand{\orcidicon}{%
	\begin{tikzpicture}
	\draw[lime, fill=lime] (0,0) 
	circle [radius=0.16] 
	node[white] {{\fontfamily{qag}\selectfont \tiny ID}};
	\draw[white, fill=white] (-0.0625,0.095) 
	circle [radius=0.007];
	\end{tikzpicture}
	\hspace{-3.5mm}
}

\foreach \x in {A, ..., Z}{%
	\expandafter\xdef\csname orcid\x\endcsname{\noexpand\href{https://orcid.org/\csname orcidauthor\x\endcsname}{\noexpand\orcidicon}}
}

\begin{document}

\preprint{APS/123-QED}

\title{Friedrich-Wintgen bound states in the continuum in dimerized dielectric metasurfaces }

\author{Xia Zhang\orcidA{}}
\email{zhangxia1@mail.neu.edu.cn}

 \affiliation{College of Sciences, Northeastern University, Shenyang 110819, China}
\author{A. Louise Bradley\orcidB{}}%
 \email{bradlel@tcd.ie}
\affiliation{%
School of Physics, CRANN and AMBER, Trinity College Dublin, Dublin, Ireland and IPIC, Tyndall National Institute, T12 R5CP Cork, Ireland.
}%

\date{\today}

             

\begin{abstract}

Bound states in the continuum (BIC) are trapped eigenmodes with infinite $Q$ factors that are confined in the system. In this work, we propose a simple design for engineering a Friedrich-Wintgen BIC through the interference between a symmetry protected BIC and a surface lattice mode in a dimerized dielectric metasurface. The meta-atoms are comprised a symmetric double bar dimer. Using incident angle tuning an avoided crossing between the symmetry protected BIC and surface lattice mode is observed. At a specific detuning, the lower energy resonance vanishes, resulting in the formation of a Friedrich-Wintgen BIC. Investigations using coupled mode theory elucidate the role of the damping rate and coupling strength in the formation of the Friedrich-Wintgen BIC in the dimerized bar metasurface. By tuning the spacing between the two bars, Friedrich-Wintgen BIC can be engineered with controlled energy and damping rate. It is shown that the damping rate of the coupled modes can be considerably suppressed in this system. We envision that these results will not only enhance the understanding of strongly coupled interactions in metasurfaces but also indicate new paths for actively and passively controlling metasurface resonators for photonic applications.

\end{abstract}

\maketitle


\section{Introduction}

Coherent phenomena, arising from the physics of coupled oscillators, manifest in various guises such as Fano resonances \cite{miroshnichenko2010fano,limonov2017fano}, electromagnetically induced transparency \cite{jenkins2013metamaterial,hajian2024quasi}, broadband extraordinary optical transmission \cite{alu2011plasmonic}, nonlinear surface lattice resonances \cite{michaeli2017nonlinear}, and strong coupling \cite{shi2014spatial}.  Metasurfaces can be engineered to support many different types of resonance modes providing an ideal platform for investigating coherent phenomena and potential for their applications.  Particularly, strongly coupled resonant states possess properties that are different to those of the bare resonator, strong coupling exhibits an avoided crossing of the dispersion of the two coupled resonators. This anti-crossing forms two distinct branches in the dispersion, an upper and a lower branch, with the minimum energy separation known as Rabi splitting \cite{askitopoulos2011bragg}. In the strong coupling regime the coherent superposition of eigenmodes at near zero energy detuning are ascribed to the formation of collective states.  Collective states open up unique possibilities for engineering the spectral response by finely tailoring the system dispersions \cite{shi2014spatial,weber2023intrinsic}. 

Over the past decade, it has been demonstrated that metasurfaces can be used to customise the field amplitude and phase of the electromagnetic radiation providing control of the light propagation, polarization as well as light-matter interaction \cite{weber2023intrinsic, hu2022multiple, chen2022can,le2024super}. Bound states in the continuum (BIC) are one of the modes that can be achieved in metasurfaces and are extensively studied due to their infinitely high Q-factor.  BIC, first discovered in quantum systems by von Neumann and Wigner \cite{friedrich1985interfering} and later proved from Maxwell’s theory \cite{marinica2008bound,ndangali2010electromagnetic}, are in principle wave solutions that are embedded in a radiative continuum and ideally non-radiative, thus with an infinitely high but inaccessible $Q$. However, the eigenmode becomes accessible when the symmetry is broken \emph{via} an oblique angle of incidence or geometric symmetry breaking \cite{koshelev2018asymmetric, liu2019high,doeleman2018experimental}, where the BIC becomes a quasi-BIC and has an accessible but still finite high $Q$ \cite{hsu2016bound,liu2019high}. Therefore, BIC have proven to be an efficient way to create trapped electromagnetic modes with near-infinite $Q$ factors. BIC metasurfaces provide enormous design freedom to maximize the interaction between light and meta-atoms, thereby bringing great potential in nonlinear processes \cite{carletti2018giant}, chiral sensing \cite{chen2022can}, and topological photonics \cite{zhen2014topological,wang2020generating}. 

BIC, typically emerge from two primary mechanisms: symmetry-protected BICs (SP-BIC) and accidental BIC \cite{hsu2016bound}. SP-BIC are characterized by a mismatch in symmetry between radiative modes and the mode profiles at $\Gamma$ point (the center of the Brillouin zone), which limits many potential applications based on angular selectivity. Friedrich-Wintergen BIC (FW-BIC) or Fabry-P{\'e}rot (FP-BIC), are accidental BIC \cite{PhysRevA.32.3231,luo2022wavy,shubin2023interacting}, arise from the destructive interference between two resonant modes, which can be modulated by specific parameters and does not rely on the symmetry operations. FP-BIC emerge when two resonators are coupled to each other and the propagating phase is tailored \cite{PhysRevB.105.245417,zhang2023supercavity}. FW-BIC are formed when two resonances belonging to the same cavity fall near to each other and their energy separation can be tuned as a function of a continuous parameter. They are typically found within the avoided crossing region at the off-$\Gamma$ point, thus enabling flexible angular selectivity, especially when structural symmetry is preserved \cite{sadrieva2019multipolar,huang2022tunable}. FW-BIC have been found in the vicinity of the avoided crossing of two dispersion bands in one cavity, such as hybrid plasmonic-photonic systems \cite{azzam2018formation,kikkawa2019polarization,joseph2021exploring}, a tailor-made loop laterally coupled to a waveguide system \cite{amrani2022friedrich} or recently in a pure photonic crystal slab \cite{le2024super}, with the designed two resonances to enable coupling and interference. 

According to FW-BIC theory, alterations in the coupling strength between the interfering modes can lead to a shift in the position of the FW-BIC within the band dispersion \cite{PhysRevA.32.3231}. The ability to effectively control the FW-BIC position is highly desirable. However, most FW-BIC are based on the engineering the resonance requiring searching for the proper geometric parameters, which can be time-consuming. In a recent report, Ref. \cite{le2024super}, control of the coupling coefficient was achieved by varying the fill-factor of periodic structures, requiring time-consuming optimization. 

Herein, we propose a facile structure in which the FW-BIC can be engineered varying only the dimer separation within the meta-atom. We consider a dielectric metasurface comprised of dimer bar meta-atoms to realize the FW-BIC and to control their position in momentum space. The FW-BIC are formed due to the interference between a SP-BIC and a surface lattice resonance (SLR) mode. These two resonances are the coupled oscillators in the system. We firstly introduced the coupled oscillator model and show the role of the coupling induced cross-damping term in the formation of the FW-BIC. Subsequently, we show the avoided cross signature of the strong coupling of the SP-BIC and SLR, with the appearance of the FW-BIC in the dispersion. Furthermore, the coupling strength can be controlled by only varying the inter-spacing between bar dimers, neither changing the filling fraction of the metasurface nor \emph{via} removing or adding parts to create the symmetry breaking. Moreover, we demonstrate that the FW-BIC can be freely designed in momentum space \emph{via} destructive interference of SP-BIC mode and SLR mode and we study the evolution of FW-BIC in momentum space. 
We expect that accurate design and control FW-BIC can pave the way towards realizing coherent active control in lasing devices.

\section{Theory}

Supposing that two resonances can spontaneously emit photons into the surrounding vacuum field modes, these photons can then be reabsorbed, resulting in an incoherent exchange of energy between both systems, without any definite phase relationship. This process can be described by a non-Hermitian matrix element, the so-called cross-damping term. Without taking into account the nonradiative damping terms, the amplitudes of two resonant modes evolve in time with the Hamiltonian as

\begin{equation}
H=
\begin{bmatrix}
E_1 & \kappa\\
\kappa & E_2
\end{bmatrix}
-i
\begin{bmatrix}
\gamma_1 & e^{i\phi}\sqrt{\gamma_1\gamma_2}\\
e^{i\phi}\sqrt{\gamma_1\gamma_2} & \gamma_2
\end{bmatrix}
\end{equation}
where $E_{1,2}$ and  $\rm \gamma_{1,2}$ represent the energy and the radiative damping rates, or the spectral width of the uncoupled resonances, respectively. $\kappa$ is the internal (near-field) coupling strength between the resonant modes. $e^{i\phi}\sqrt{\gamma_1\gamma_2}$ is the via-the-continuum coupling term since the two resonances interfere and radiate into the same channel, where $\phi$ is the relative dephasing for the inspected two modes radiating into the continuum. $e^{i\phi}=1$ due to the in-plane coupling studied here. By solving $\rm |H-EI|=0$, where $\rm I$ is the identity matrix, the derived eigenvalues of Hamiltonian are as follows
\begin{equation}
\label{Edis}
\begin{split}
   \rm  E_{\pm}&= \rm \frac{E_1+E_2}{2}-i\frac{\gamma_1+\gamma_2}{2}\\&
   \rm  \pm \frac{1}{2} \sqrt{[(E_1-E_2-i(\gamma_1-\gamma_2)]^2+4(\kappa-i\sqrt{\gamma_1\gamma_2})^2}
\end{split}
\end{equation}

Hybridization of the two uncoupled modes leads to the formation of two branches of hybrid (dressed) modes with a clear splitting, or displaying an avoided crossing behavior. The minimum energy spacing between the two branches, referred as the vacuum Rabi splitting, becomes

\begin{equation}
\label{Omega_R}
\begin{split}
\Omega_R&=Re|E_+-E_-|_{min}\\&
=\sqrt{[(E_1-E_2-i(\gamma_1-\gamma_2)]^2+4(\kappa-i\sqrt{\gamma_1\gamma_2})^2}
\end{split}
\end{equation}
When the condition $\rm E_1=E_2$ is satisfied, or at the zero energy detuning case, it yields $\rm \rm \Omega_R= \sqrt{4\kappa^2-(\gamma_1+\gamma_2)^2}$. To guarantee coherent and reversible energy transfer, the energy anti-crossing behavior can only be resolved when the Rabi splitting is larger than the total dissipation energy of the hybrid system, which can be expressed as 
\begin{equation}
\label{SCcriteria}
\rm \kappa> \frac{\gamma_1+\gamma_2}{2}
\end{equation}
A special case exists when the conditions $\rm E_1=E_2$ and $\rm \gamma_1=\gamma_2$ are simultaneously met, and the two modes are in-phase. The eigenvalues become
\begin{equation}
\begin{split}
&\rm E_{+}=E_1+ \kappa - 2i \gamma \\&
\rm E_{-}=E_1-\kappa
\end{split}
\end{equation}
implying resonance energy splitting and linewidth changes. The derived Rabi splitting energy becomes, $\rm \Omega_R = 2\kappa$. The criterion of strong coupling according to Eq.~\ref{SCcriteria} becomes $\rm \kappa>\gamma_1=\gamma_2$. It is clear that the Rabi splitting and coupling strength depend critically on the energy detuning and the damping rates of the uncoupled resonances. 

The other special case is for the FW-BIC condition when 
\begin{equation}
\label{FWcriteria}
\kappa(\gamma_1-\gamma_2)=\sqrt{\gamma_1\gamma_2}(E_1-E_2)
\end{equation}
with the corresponding eigenvalues, upper band, $E_+$ and lower band, $E_-$, expressed as 
\begin{equation}
\label{FWeigenvalues}
\begin{split}
&E_+=\frac{E_1+E_2}{2}+\frac{\kappa(\gamma_1+\gamma_2)}{2\sqrt{\gamma_1\gamma_2}}-i(\gamma_1+\gamma_2)\\&
E_-=\frac{E_1+E_2}{2}-\frac{\kappa(\gamma_1+\gamma_2)}{2\sqrt{\gamma_1\gamma_2}}
\end{split}
\end{equation}
It is evident that at the FW-BIC position, the imaginary part of the eigenvalue $E_+$
is $\gamma_1+\gamma_2$, implying a wider spectral linewidth of the resonance, while that of $E_-$ vanishes with zero spectral linewidth, implying formation of a bound state. 
 Furthermore, the magnitude of the splitting at the FW-BIC position is 
\begin{equation}
\label{FWDelta}
\Delta=Re|E_+-E_-|= \frac{\kappa(\gamma_1+\gamma_2)}{\sqrt{\gamma_1\gamma_2}}
\end{equation}
It indicates that the energy difference at the FW-BIC position, relies on a cooperative response of coupling strength, $\kappa$ and the damping rates of the uncoupled resonances, $\gamma_1$ and $\gamma_2$.

\begin{figure}[htbp]
\includegraphics[width=1\linewidth]{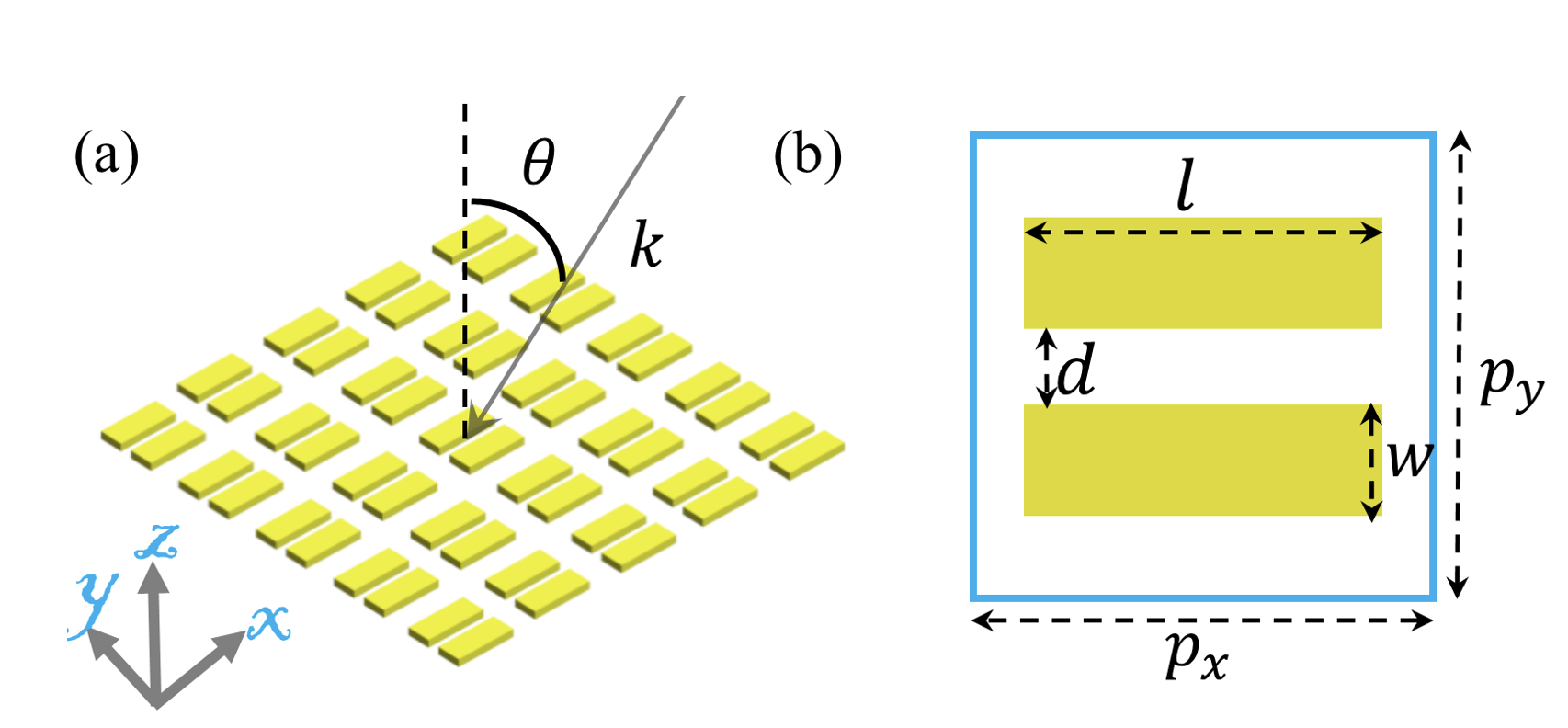}
\caption{\label{sch} Schematic graph of dielectric dimer metasurface (refractive index, $n$ = 3.6). The dimer is composed of two identical rectangular bars, with a separation, $d$. The dimers are arranged in a square lattice with a period of $p_x = p_y = 560$ nm, the length of the  rectangular bar, $l$ = 500 nm, the width $w$ = 160 nm and the thickness, $t$ = 50 nm. The metasurface is investigated in free space and is illuminated by a TE-polarized plane wave with an incident angle, $\theta$.  }
\end{figure}

\begin{figure*}[htbp]
\includegraphics[width=1\linewidth]{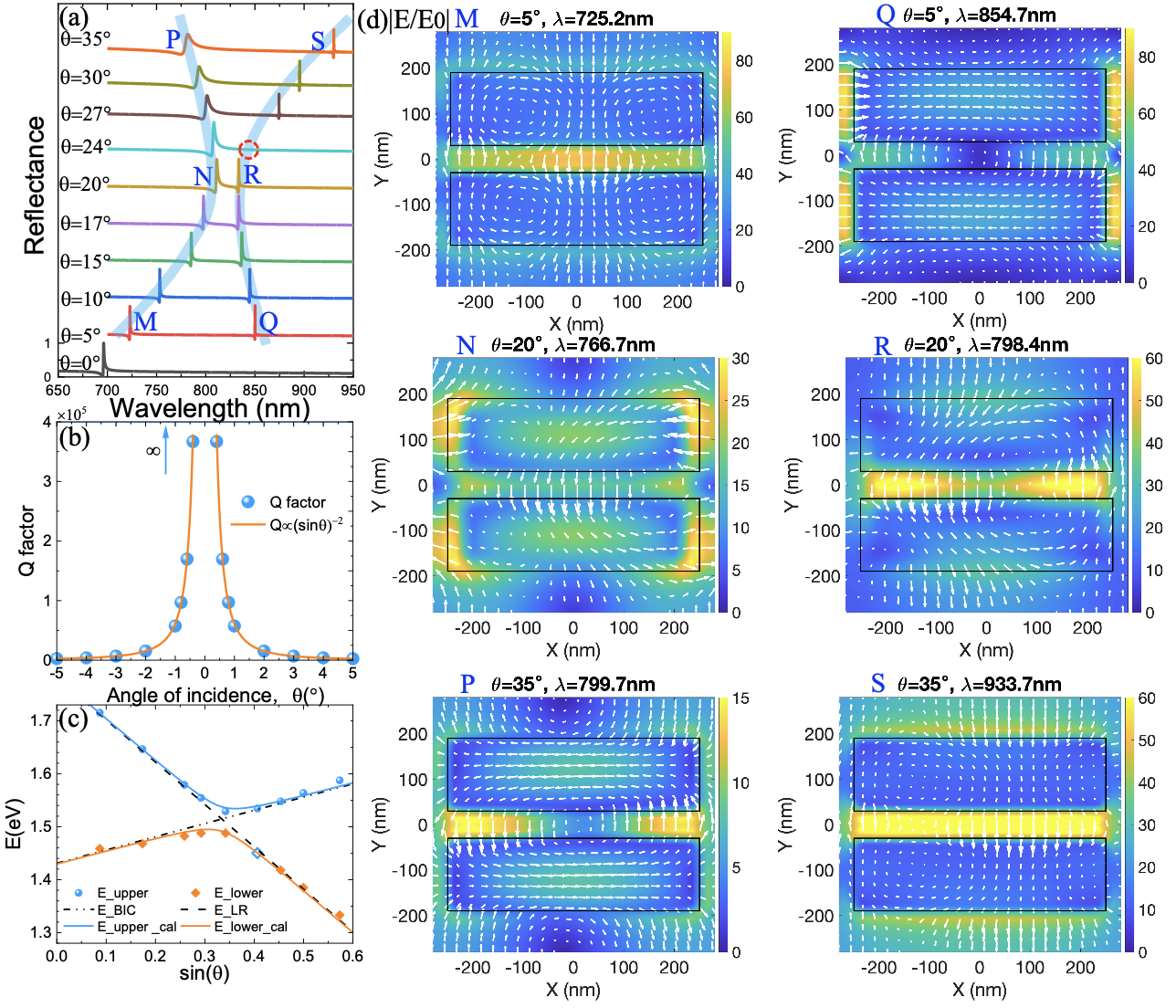}
\caption{\label{d60} (a) Stimulated reflectance spectra $\textit{vs}$ the angle of incidence, $\theta$. The shaded area illustrates the avoided crossing signature due to strong coupling between two modes, the uncoupled resonances are denoted as M for the SLR and Q for the SP-BIC mode.  The red dashed circle indicates the FW-BIC position with vanishing linewidth due to the hybridization of the SP-BIC mode and the lattice mode. (b) $Q$ factors of the SP-BIC mode,  with different angle of incidence, $\theta$. The solid line illustrates the fitted curve as $Q=Q_0(sin\theta)^{-2}$. (c) 
Relative amplitude of the electric field distributions $|E/E_0|$ in the $x-y$ plane through the middle of the dimer metasurface, where $E_0$ is the amplitude of the incident wave. The arrows denote the electric field vectors. The inspected wavelength at different angle is shown, the corresponding reflectance peaks are labelled as M, N, P, Q, R, S respectively.}
\end{figure*}

\begin{figure}[htbp]
\includegraphics[width=0.8\linewidth]{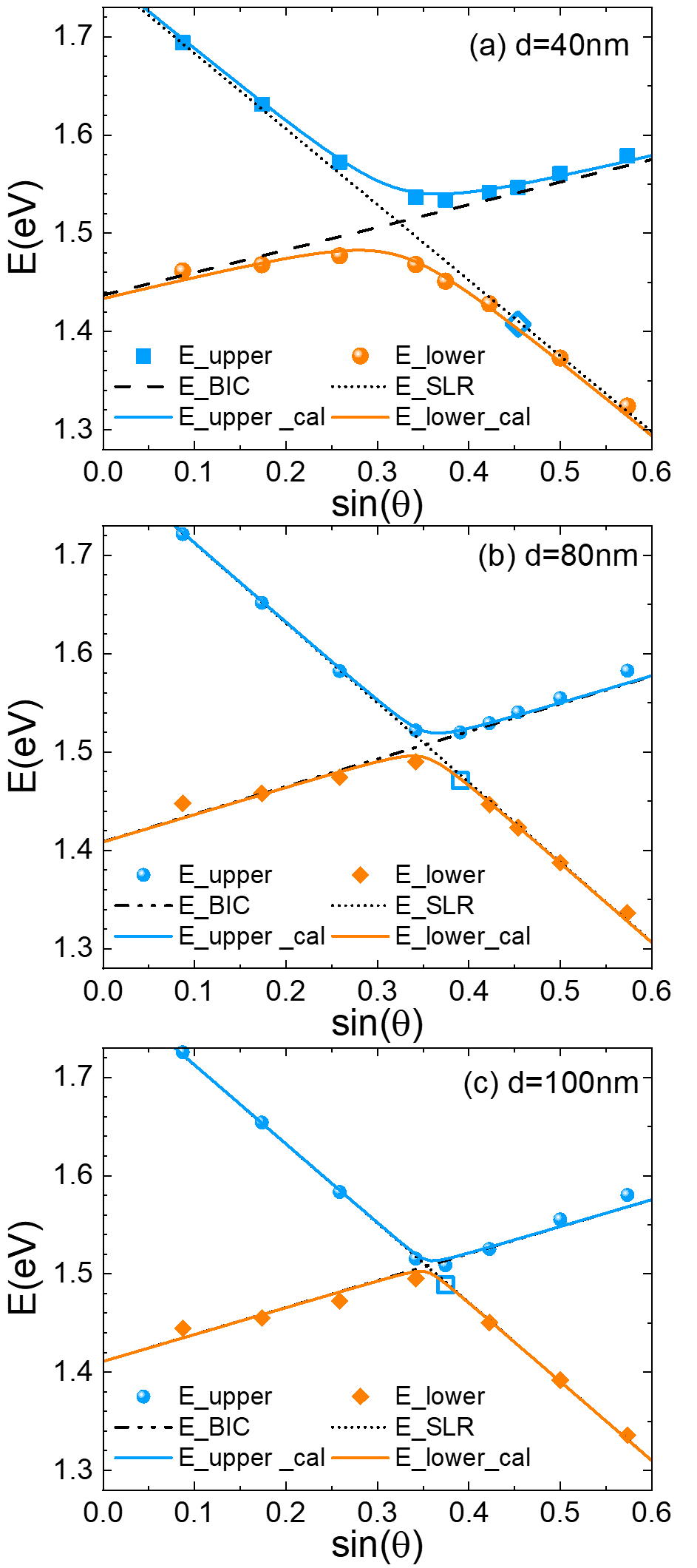}
\caption{\label{ddep} Peak reflectance ($y$-polarized incident light) energies as a function of $sin\theta$ for the dimer bar metasurace with varying distance $d$ for (a) $d$ = 40 nm, (b) $d$ = 80 nm, and (c) $d$ = 100 nm, respectively. The black dashed lines show the dispersion of uncoupled SP-BIC mode and SLR modes. Solid lines correspond to the calculated curves by Eq.~\ref{Edis}. The empty squares denote the FW-BIC position with vanishing linewidth at the lower energy branch. }
\end{figure}

\begin{figure}[htbp]
\includegraphics[width=0.75\linewidth]{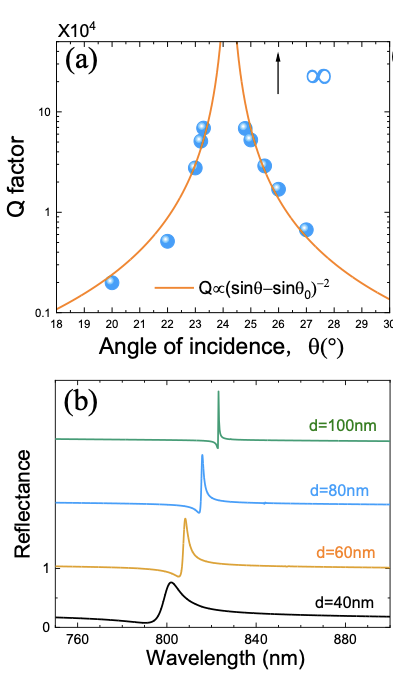}
\caption{\label{FW} (a) The $Q$ factors of the FW-BIC mode for  $d$ = 60 nm with varying angle of incidence, $\theta$. The solid line shows the fit to the data points using $Q=Q_0(sin[\theta-\theta'])^{-2}$ and $\theta'=24^\circ$. (b) The reflectance spectra of dimer bar metasurface at the FW-BIC angle for different bar separations, $d$ = 40 nm, 60 nm, 80 nm and 100 nm respectively.  }
\end{figure}

\section{Results and Discussion}

\subsection{Strong coupling of SP-BIC and SLR modes}

The geometry of the metasurface is sketched in Fig.~\ref{sch} (a), which consists of a two-dimensional array of dimer rectangular bars in air. The dimer rectangular bars are separated by a distance, $d$ = 60 nm, as seen in Fig.~\ref{sch} (b), the height of the bar, $h$ = 50 nm, and the period, $p_x=p_y=560$ nm, are kept fixed. The refractive index of the bar is set as $n$ = 3.6 (corresponding to Si in the near-infrared wavelength range) and the background medium is in air. The metasurface is normally and obliquely illuminated by a plane wave with unitary electric field amplitude ($|E_0|$= 1) and $y$ polarization.  The reflectance spectra and electric field maps of the dimer rectangular bar metasurface were calculated by a finite-difference time-domain method (Ansys Lumerical FDTD). For normal light incidence, the Bloch/periodic plane waves were employed, and for the oblique light incidence, the broadband fixed angle source technique (BFAST) source was implemented.

Fig.~\ref{d60} (a) presents the simulated reflectance spectra of the rectangular bar dimer metasurface under varying angles of incidence, $\theta$ for $y$-polarized light. Two pronounced modes: a SLR mode and a SP-BIC mode, can be observed. The SLR mode, such as M depicted at $\theta$ = 5$^\circ$ red shifts with increasing angle of incidence. The SP-BIC mode, such as Q at $\theta$ = 5$^\circ$, blue shifts with increasing angle, approaching the SLR mode gradually, guaranteeing the coupling of the two modes. Both modes approach each other at around $\theta$ = 20$^\circ$, displaying clear avoided crossing signatures.

It is seen that SP-BIC mode is absent at $\theta$ = 0$^\circ$ but appears under oblique angle of incidence due to breaking the in-plane mirror symmetry of the structure. To further unveil the physical origin of the quasi-BIC mode, the quality factor, $Q$ evolution as a function of incident angle, $\theta$ is shown in Fig.~\ref{d60} (b). Oblique angles of incidence, behaving as small perturbations, are introduced into the resonant system, and the SP-BIC will transformation into a quasi SP-BIC, exhibiting a high-$Q$ resonance and tight optical field confinement \cite{koshelev2018asymmetric}. The resonances appearing in the simulated reflectance spectra display Fano profiles and can be phenomenologically fitted by $T(E)=1-R(E)=T_0+A_0\frac{[q+2(E-E_0)/\Gamma]^2}{1+[2(E-E_0)/\Gamma]^2}$ \cite{PhysRevA.57.4407,miroshnichenko2010fano},
where $E_0$ is the resonance energy and $E = hc/\lambda$; $\Gamma$ is the resonance energy linewidth, or full width at half maximum; $\lambda$ is the free space wavelength; $T_0$ is the transmission offset; $A_0$ is the continuum-discrete coupling constant, and $q$ is the Breit-Wigner-Fano parameter determining the asymmetry of the resonance profile. The $Q$ factor of the Fano resonance is evaluated by $Q=E_0/\Gamma$. As can be seen in Fig.~\ref{d60} (b), the calculated $Q$ factors obey the inverse-square law that $Q\propto(sin\theta)^{-2}$, consistent with Ref. \cite{koshelev2018asymmetric}, a demonstration of quasi-BIC nature. Moreover, the unique polarization vortex center at the BIC modes can serve as further evidence of the BIC nature \cite{zhen2014topological}. As can be seen in Fig.~\ref{d60} (d) for $\theta=5^{\circ}$ at $\lambda=854.7$ nm, a closed loop electric field profile in the $x-y$ plane through the middle of the dimer metasurface is seen with counterclockwise direction, which is another typical signature of quasi-BIC.  

Furthermore, Fig.~\ref{d60} (d) depicts the amplitude of electric field ratio, $|E/E_0|$ in the $x-y$ plane through the middle of the dimer metasurface, corresponding to six different wavelengths illustrated in Fig.~\ref{d60} (a), which are labelled M, N, P, Q, R and S respectively. It is seen that Q and P display similar vortex electric field patterns, indication of SP-BIC evolution as a function of incident angle. With increasing incident angles, the amplitude of the electric field decreases gradually due to increasing asymmetry factor, corresponds well with decreasing $Q$ factors. Furthermore, M and S show similar pattern, displaying the evolution of the SLR mode. It is important to emphasize that in the avoided crossing region, the electric field amplitude ratio, $|E/E_0|$ at N and R, is determined by the superposition of the electric fields of the SLR mode and the SP-BIC mode. Generally, SLR modes are generated through the diffractive coupling of incident light with the modes sustained by the array \cite{PhysRevB.25.689,michaeli2017nonlinear,PhysRevB.54.6227}. The in-plane or tangential component of the wave vector in the metasurface can be described by 
$\vec{k}_x=\vec{k}_{x\theta_{in}}\pm\vec{G}(m_1,m_2)$ and ${k}_{x\theta_{in}} = k_0sin \theta$, where $\vec{k}_0$ is the wave vector of the incident beam, and $\vec{G}(m_1,m_2)=m_1({2\pi}/{a_x})\vec{x}+m_2({2\pi}/{a_y})\vec{y}$ is the reciprocal lattice vector of the array with $m_1$ and $m_2$ the orders of diffraction in the $x$ and $y$ directions, respectively. For the oblique angle of incidence with the polarization along $y$ direction, the reflectance peaks shift spectrally following the Rayleigh anomaly conditions. The in-plane wave vector of $\vec{k}_x$ is parallel to the $x$ axis of the sample with magnitude $k_{//}=k_0 sin\theta=2\pi/\lambda sin\theta$, implying approximate linear resonance shift with $sin\theta$ of the uncoupled SLR mode.

In order to qualitatively describe the concurrence of avoided crossing of SLR and SP-BIC modes, coupled mode theory, Eq.~\ref{Edis} is employed to interpret the data, shown as the solid lines in Fig.~\ref{d60} (c). The energy data points are extracted from  the reflectance peaks in (a) and plotted as a function of sin$\theta$. The dispersion relation is correspondingly constructed, displaying the characteristic avoided crossing behaviour of the lower and upper branches. When the energies of the uncoupled SP-BIC mode and SLR mode are equal (zero detuning), the Rabi splitting energy can be extracted as the minimal energy splitting between the upper and lower branches, $\Omega_R$ = 44.9 meV for $d$ = 60 nm, according to Eq.~\ref{Omega_R}.  Furthermore, the dispersion of uncoupled SP-BIC (denoted as mode 1) and SLR modes (denoted as mode 2) can be approximated as $E_{1}=\alpha_{1}sin\theta+E_{1C}$ and $E_{2}=\beta_{2}sin\theta+E_{2C}$, respectively. Moreover, Eq.~\ref{Edis} needs to consider the damping rates of the uncoupled modes, $\gamma_{1}$ for SP-BIC mode and $\gamma_{2}$ for SLR mode, respectively. The radiative losses of SLR modes are assumed to be weakly dependent on the incident angle and is thus treated as a constant  \cite{heilmann2020strong,peng2020separation,le2024super}.  The damping rate of the SLR mode, $\gamma_{2}$ for $d$ = 60 nm was calculated to be 0.72 meV, defined by the Fano fit of the reflectance spectrum at incident angle of 5$^\circ$, or M illustrated in Fig.~\ref{d60} (a). It is further verified that the damping rate of the uncoupled SLR mode is the same for the incident angles of 5$^\circ$, 10$^\circ$ and 15$^\circ$. Meanwhile, due to angular dependence of the $Q$ factors as illustrated in Fig.~\ref{d60} (b) and slight resonant wavelength shift, the damping rates of of the SP-BIC modes are estimated as $\gamma_{1}=\gamma_0 sin^2\theta=0.015\times sin^2\theta$, where $\gamma_0$ is estimated from the Fano fit of the uncoupled SP-BIC modes, or reflectance peaks at incident angle of 5$^\circ$, 10$^\circ$ and 15$^\circ$ respectively. Therefore, the coupling parameter, coupling strength, $\kappa$ becomes the only fit parameter. As can be seen from the analytical curves shown in Fig.~\ref{d60}  (c), a coupling strength, $\kappa = 23$ meV, agrees well with the numerically simulated data, demonstrating the validity of our approach. It can be noted that the damping rate of the SP-BIC at the Rabi splitting position ($\theta$ = 20$^\circ$) is $\gamma_{1}=\gamma_0sin^2\theta=0.015\times sin^2 20^\circ$ = 1.75 meV. Together with the estimated damping rate of the SLR mode at the Rabi splitting position, $\gamma_{2}$ = 0.072 meV, it is clear that $\kappa> (\gamma_1+\gamma_2)/2$, consistent with the criterion for strong coupling regime of the SLR mode and SP-BIC modes. 

\begin{table}[htbp]
\caption{ Extracted physical parameters for dimer rectangular bar metasurfaces with various bar separation, $d$ = 40 nm, 60 nm, 80 nm and 100 nm respectively, including Rabi splitting energy: $\Omega_R$, coupling strength: $\kappa$, Rabi splitting angle: $\theta$,
FW-BIC angle: $\theta'$,  radiative damping rate of the lossy branch at the FW-BIC angle: $\gamma_1+\gamma_2$, energy splitting at the FW-BIC angle: $\Delta$,  the fitting parameters for the uncoupled SP-BIC mode, denoted as mode 1: $E_{1C}$, $\alpha_1$ and that for SLR mode, denoted as mode 2: $E_{2C}$, $\beta_2$. }
\begin{tabular}{ |p{3.4cm}|p{1.1cm}|p{1.1cm}|p{1.1cm} |p{1.1cm}|}
\hline
 \hline
$d$(nm) & 40 &  60 & 80 & 100  \\ 
  \hline
  $\Omega_R$(meV)   & 67.5  & 44.9    &  29.6 & 21.3\\
  $\kappa$(meV)   & 34  & 23    &  13.2 & 3 \\
  Rabi angle $\theta$($^{\circ}$)  & 20 & 20   & 20 &  20  \\
   \hline
  FW-BIC angle $\theta'$ ($^{\circ}$)  & 27 & 24   & 23 &  22  \\
 FW-BIC $\gamma_1+\gamma_2$ (meV)  & 16.6   & 4.2     & 2.3
 & 1 \\
  FW-BIC $\Delta$(meV)   & 140  &   86.5  & 47.5  & 20 \\
   \hline
$E_{1C}$(eV)   & 1.437   & 1.431      & 1.409  & 1.411 \\
 $\alpha_1$ (eV)    & 0.23  & 0.25     & 0.28   & 0.274 \\
  $E_{2C}$ (eV)  & 1.76   & 1.782     & 1.793    & 1.793 \\
 $\beta_2$ (eV)   &  -0.77  & -0.8    & -0.810  & -0.809\\
\hline
  \hline
\end{tabular}
\label{table1}
\end{table}

\subsection{FW-BIC formation }

Particular attention is paid to $\theta=24^\circ$ in Fig.~\ref{d60} (a), illustrated by a red dashed circle or the empty square as seen in Fig.~\ref{ddep} (c). In the reflectance spectra in Fig.~\ref{d60} (a), each mode can be described by an asymmetric Fano line shape. Vanishing of the Fano lineshape occurs on the lower energy branch at $\theta=24^\circ$ with $\gamma_{-}= 0$, indicating that the BIC emerged with an infinite $Q$-factor via the destructive interference of the SP-BIC mode and SLR mode. Whereas, the resonance in the upper energy branch becomes more lossy as $\gamma_{+}=\gamma_{1}+\gamma_{2} = 4.2$ meV.  Furthermore, the energy spacing at the FW-BIC position is $\Delta=Re|E_+-E_-|$ = 86.5 meV, which is located in the vicinity of the Rabi splitting regime.  The phenomenon meets the FW-BIC criterion in Eq.~\ref{FWcriteria} and Eq.~\ref{FWeigenvalues}. The validation of the FW-BIC nature can be further accomplished by its $Q$ factor evolution, as calculated for different angles and depicted in Fig.~\ref{FW}(a). It is evident that the $Q$ factor increases as the incident angle approaches a specific angle at $\theta=24^\circ$, with infinite $Q$ factor, indicating the formation of a true BIC mode. When the incident angle $\theta$ continues to increase, the reflectance peak appears again, indicating the transformation from a BIC to quasi-BIC mode. The $Q$-factor dependence can be well described by the formula $Q_0 (sin\theta-sin\theta')^{-2}$ \cite{hu2022multiple}, where $\theta' =24^\circ$ corresponds to the FW-BIC angle. 

As discussed earlier, the damping rate of the SP-BIC modes at varying angle of incidence, $\theta$ are estimated as $\gamma_{1}=\gamma_0 sin^2\theta=0.015\times sin^2\theta$ and the damping rate of the SLR mode is considered constant. It is clear that: the total damping rate of both modes at the FW-BIC angle is larger than at the Rabi splitting angle (described as $\gamma_{10}$,$\gamma_{20}$): $\gamma_{1}+\gamma_{2}>\gamma_{10}+\gamma_{20}$. Therefore, if the condition, $\kappa>(\gamma_{1}+\gamma_{2})/2$, at the FW-BIC angle holds, we can conclude: $\kappa>(\gamma_{10}+\gamma_{20})/2$, with evidence of strong coupling between the SLR and the SP-BIC resonances.

In order to further investigate the SLR and SP-BIC mode coupling behavior as well as the formation of a FW-BIC mode, dispersion curves of dimer bar metasurfaces with various separation of the bars, $d$ are explored. The dispersion for $d$ = 40 nm, 80 nm and 100 nm are seen in Fig.~\ref{ddep}. An obvious avoided crossing behavior is exhibited for all cases. The extracted Rabi splitting energy, $\Omega_R$, the coupling strength, $\kappa$, the angle where FW-BIC appears, $\theta'$, the damping rate of the lossy upper branch at the FW-BIC angle, $\gamma_1+\gamma_2$, the energy splitting at FW-BIC position, $\Delta$, as well as the fitting parameters for the uncoupled SP-BIC mode, denoted as mode 1: $E_{1C}$, $\beta_1$ and that for SLR mode, denoted as mode 2: $E_{2C}$, $\alpha_2$ are all summarized in Table~\ref{table1}.  

The FW-BIC were observed with vanishing peaks and linewidths, near the avoided crossing point for all cases including $d$ = 40 nm, 80 nm and 100 nm, respectively, illustrated by empty squares.  With vanishing linewidths observed in the lower energy branch at the FW-BIC angle, the energy linewidth of the higher energy branch become more lossy as $\gamma=\gamma_{1}+\gamma_{2}$, and can be seen in the reflectance spectra in Fig.~\ref{FW} (b).  The $\gamma_{1}+\gamma_{2}$ values for varying bar separation, extracted at the FW-BIC angle from the reflectance spectra in Fig.~\ref{FW} (b), are given in Table~\ref{table1}. They are found to be 16.6 meV, 4.2 meV, 2.3 meV and 1 meV for $d$ = 40 nm, 80 nm and 100 nm, respectively. It is clear that $\kappa>(\gamma_{1}+\gamma_{2})/2>(\gamma_{10}+\gamma_{20})/2$, therefore, SP-BIC mode strongly couples with the SLR mode in all cases. It is a well-known feature that at the FW-BIC position, the damping rate of the lower branch vanishes exactly due to destructive interference of SP-BIC mode and SLR modes, leading to infinite $Q$ factors while the other is boosted to maximum due to constructive interference. However, it is found that the damping rate of upper branch at FW-BIC position, $\gamma_1+\gamma_2$, can also be suppressed $\textit{via}$ only increasing the bar separation, $d$. The radiation suppression is attributed to the elimination of the first Fourier harmonic component in the lattice parameters studied elsewhere \cite{gao2022dark,lee2021metasurfaces}. 

Rabi splitting, with minimum energy spacing of the two branches, occurs at approximately 20$^{\circ}$. However, the FW-BIC position occurs at 27$^{\circ}$ for $d$ = 40 nm,  24$^{\circ}$ for $d$ = 60 nm, 23$^{\circ}$ for $d$ = 80 nm
and 22$^{\circ}$ for $d$ = 100 nm, further shifting to the avoided crossing point with increasing inter-separation, $d$. Correspondingly, the energy difference at the FW-BIC, $\Delta=E_+-E_-$, decreases with bar separation, $d$ and approaches gradually the Rabi splitting energy, $\Omega_R$. These observations agree with eq.~\ref{FWDelta}, $\Delta=Re|E_+-E_-|= {\kappa(\gamma_1+\gamma_2)}/{\sqrt{\gamma_1\gamma_2}}$, where the position of the FW-BIC relies on the coupling strength, $\kappa$, as well as the damping rate of the uncoupled resonance at the FW-BIC position. As the separation $d$ increases, $\kappa$ and $\gamma_{1}+\gamma_{2}$
decrease, which in turn affects the position of the FW-BIC. All these parameters collectively determine the interference conditions necessary for the formation of FW-BICs.


\section{Summary}

In summary, we explore the interaction between a SP-BIC mode and a SLR mode in a dimerized dielectric bar metasurface. Through both analytical and numerical methods, we demonstrate the formation of a FW-BIC near an avoided crossing, resulting from the interference between these modes. We differentiate the SP-BIC and SLR modes based on their distinct electric field patterns and dispersion behaviors, which vary with the angle of incidence. Using temporal coupled-mode theory, we provide insights into the strong coupling mechanism and describe how an accidental FW-BIC emerges in this system.
Our findings reveal that the FW-BIC is loosely associated with the avoided crossing point. Its position can be adjusted by changing the separation between the dimer bars, allowing for control over the coupling strength without altering the metasurface's geometry design.
These results have significant implications for the development of FW-BIC-based technologies, such as filters, sensors, nonlinear optical devices, and low-threshold lasers. We anticipate that our work will advance the engineering of FW-BICs and inspire their application in various photonic devices.

\nocite{*}
\begin{acknowledgments}
We wish to acknowledge the support of the Fundamental Research Funds for the Central Universities of Ministry of Education of China (Grant No. N2405012) and Science Foundation Ireland Frontiers for the Future Award (Grant No. SFI-21/FFP-P/10187).
\end{acknowledgments}

\bibliography{apssamp}

\end{document}